\documentclass[prb,floatfix,twocolumn,showpacs,amsmath,amssymb]{revtex4}
\usepackage{graphicx,color}
\usepackage{dcolumn}
\usepackage{bm}
\begin{document}

\title{Suppression of Dimer Correlations in the Two-Dimensional $J_1$-$J_2$ Heisenberg Model: an Exact Diagonalization Study}
\author{Luca Capriotti,$^{1}$ Federico Becca,$^{2,4}$  Alberto Parola,$^{3}$ and Sandro Sorella$^{4}$}
\affiliation{
${^1}$  Kavli Institute for Theoretical Physics, University of California, Santa Barbara CA 93106-4030 \\
${^2}$  Institut de Physique Th\'eorique, Universit\'e de Lausanne, CH-1015 Lausanne, Switzerland \\
${^3}$  Istituto Nazionale per la Fisica della Materia and Dipartimento di Scienze, Universit\`a dell'Insubria, I-22100 Como, Italy \\
${^4}$ INFM-Democritos, National Simulation Centre, and SISSA, I-34014 Trieste, Italy.}
\date{\today}

\begin{abstract}
We present an exact diagonalization study of the ground state of the 
spin-half $J_1{-}J_2$ model.
Dimer correlation functions and the susceptibility associated to the breaking
of the translational invariance are calculated for the $4\times 4$ and the 
$6\times 6$ clusters. These results -- especially when compared to the
one dimensional case, where the occurrence of a dimerized phase for large
enough frustration is well established  -- suggest either a homogeneous
spin liquid or, possibly, a dimerized state with a rather small 
order parameter. 
\end{abstract}

\maketitle

In quantum systems, frustration, i.e., the impossibility
to satisfy simultaneously every pairwise interaction, induces
unconventional properties down to zero temperature ($T=0$).
In particular it plays a very important role in quantum antiferromagnets
where the combined effect of competing interactions and zero-point motion
can give rise to non-magnetic 
ground states which lack the classical N\'eel long-range order.
To this respect the most intriguing situation appears 
in two spatial dimensions where at $T=0$
both the magnetic and non-magnetic ground states are allowed by 
the Mermin-Wagner theorem, and quantum phase transitions from an ordered to 
a disordered magnetic phase can be triggered by increasing the 
frustration.~\cite{sachdev1}

One of the simplest frustrated spin systems in two dimensions is the 
so-called $J_1{-}J_2$ Heisenberg model.~\cite{doucot}
Here the antiferromagnetic coupling ($J_1>0$) between nearest 
neighbor ({\it n.n.}) spins is frustrated by the presence of a next-nearest neighbor 
({\it n.n.n.}) superexchange interaction ($J_2>0$):
\begin{equation} \label{j1j2ham}
\hat{\cal{H}}=J{_1}\sum_{n.n.}
\hat{{\bf {S}}}_{i} \cdot \hat{{\bf {S}}}_{j}
+ J{_2}\sum_{n.n.n.}
\hat{{\bf {S}}}_{i} \cdot \hat{{\bf {S}}}_{j}~,
\end{equation}
where $\hat{{\bf {S}}}_{i}$ are spin-half operators on a periodic
$N-$site square lattice.
Recently, the interest in this model has been boosted by the synthesis
of three vanadates compounds (${\rm Li_2VOSiO_4}$, ${\rm Li_2VOGeO_4}$, and
${\rm VOMoO_4}$)
whose relevant magnetic interactions involve nearest and next-nearest
spin-$1/2$ $V^{4+}$ ions on weakly coupled stacked planes.~\cite{carretta}

Despite the simplicity of the Hamiltonian (\ref{j1j2ham}),
the zero-temperature phase diagram of the spin-half $J_1{-}J_2$ model
has been much debated in the last 15 years.
For $J_2/J_1 \ll 0.5$,
an antiferromagnetic (AF) N\'eel order is expected, 
while in the opposite limit, 
$J_2/J_1 \gg 0.5$ the ground state should be in an AF collinear phase, 
with the spins ferromagnetically aligned in one direction and antiferromagnetically
in the other, corresponding to magnetic wave vectors
$Q=(\pi,0)$ or $Q=(0,\pi)$.
The most interesting
part of the phase diagram is the one around the fully frustrated point
$J_2/J_1=0.5$. In fact, although the existence of a gapped (non-magnetic)
phase is at present rather likely for 
$0.4\lesssim J_2/J_1\lesssim 0.55$,~\cite{doucot,luca,russi} 
its characterization is one
of the most intriguing puzzles of the physics of strongly correlated systems.
In particular, a debated issue is whether the ground state is a 
homogeneous spin liquid, as originally proposed by 
Anderson and Fazekas.~\cite{fazekas,figuerido} The other possibility, 
following Read and Sachdev's large-$N$ expansion,~\cite{read} is a 
spontaneously dimerized phase which breaks the translation invariance of the
Hamiltonian or -- more in general -- one of the lattice symmetries.~\cite{russi,singh,russi2,leuwen}

The main reason for the lack of agreement on this delicate issue
is due to the fact that most of the existing results
have been obtained with approximated numerical and analytical techniques 
based on reference states explicitly breaking some symmetry of the 
Hamiltonian,~\cite{luca,russi,singh,russi2,leuwen}
so that it is very difficult to put on a solid ground any conclusion on the 
actual ground-state correlations. 
In this context, the unbiased results provided by the
exact diagonalization (ED) of small clusters can give
a valuable insight into the $T=0$ properties of the model.~\cite{poilblanc} 
In particular, on the $4 \times 4$ cluster, Poilblanc and 
collaborators~\cite{poilblanc2} showed that the dimer correlations are
enhanced in the parameter range $0.5<J_2/J_1<0.6$, concluding that the
dimer state is a good candidate for describing the physics of the
$J_1{-}J_2$ model in the highly-frustrated region. 
This claim is questionable because of the smallness of the
$4 \times 4$ lattice, where it is very difficult to extract the important
long-range behavior of the correlation functions.
In this paper, we enlarge the cluster up to $N=36$ sites and present a
more systematic analysis based on {\it i}) the calculation of the dimer 
susceptibility in the presence of an explicit symmetry-breaking term and 
{\it ii}) the calculation of a static dimer correlation function.
In both cases, the direct comparison with the one-dimensional chain, 
where the dimerization is well established for large enough
frustration,~\cite{emery1d} casts doubts on the existence of a dimerized 
two-dimensional ground state in the non-magnetic regime. 

The occurrence of some kind of dimerized order can be detected by calculating 
the response of the system to operators breaking the lattice symmetries. 
This can be done by adding to the Hamiltonian
a term $-\delta \hat{O}$, where $\hat{O}=\sum_i\hat{O}_i$
is an (extensive) hermitian operator that breaks some symmetry of $\hat{\cal H}$ and $\delta$
is a (small) parameter.
In fact, if true long-range order in the dimer correlations exists in the 
thermodynamic ground state, the finite-size susceptibility 
$\chi_{O} = 2 \langle\psi_0|\hat{O}(E_0-\hat{\cal H})^{-1}\hat{O}|\psi_0\rangle/NJ_1$
has to diverge with the system size. 
In particular, it can be shown that it is bounded from below by the 
system volume squared, $\chi_O > {\rm const} \times p^4 N^2$,
where $p=\sqrt{\langle\psi_0|\hat{O}^2|\psi_0\rangle/N^2}$ is the order parameter.~\cite{st,caprio}
A spontaneously broken translation symmetry 
can be detected by studying the
response of the system to the perturbation $\delta \hat{O}_{\rm T}$, with
\begin{equation}\label{trasla}
\hat{O}_{\rm T}=\sum_{j} e^{i Q \cdot R_j}
\hat{{\bf {S}}}_{j} \cdot \hat{{\bf {S}}}_{j+x}~,
\end{equation}
where $R_j$ indicates the coordinates of the site $j$ and $x=(1,0)$.
This operator preserves the SU(2) symmetry of the $J_1{-}J_2$ Hamiltonian
but breaks the translational invariance with momentum $Q=(\pi,0)$
and the $x \leftrightarrow y$ symmetry (or equivalently the $\pi/2$
rotation symmetry).

\begin{figure}
\includegraphics[width=0.45\textwidth]{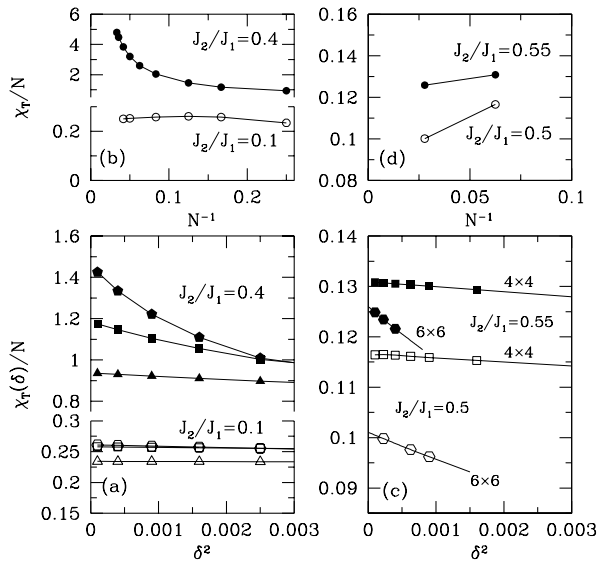}
\caption{\label{pig}
ED calculation of the susceptibility associated to the
operator $\hat{O}_{\rm T}$ of Eq.~(\protect\ref{trasla}).
Left panels: results for the one-dimensional chain.
Bottom: $\chi_{\rm T}(\delta)/N$ vs $\delta^2$
for (from below) $4$, $6$, and $8$ sites. $J_2/J_1=0.1$ (empty symbols),
0.4 (full symbols).
Top: size scaling of $\chi_{\rm T}/N$. Right panels: same quantities for the
two-dimensional case for
$J_2/J_1=0.5$ (empty symbols) and 0.55 (full symbols).
}
\end{figure}

By using a numerical technique, like the Lanczos method, the susceptibility
$\chi_{\rm T} = -d^2e(\delta)/d\delta^2|_{\delta=0}$ 
can be calculated with only energy measurements by computing the 
ground-state energy per site (in unit of $J_1$) in presence
of the perturbation, $e(\delta)$, for few values of $\delta$ and by estimating numerically
the limit
$
\chi_{\rm T}= \lim_{\delta \to 0} \chi_{\rm T}(\delta)=-{2(e(\delta)-e_0)}/{\delta^2}~.
$
Notice that, in the presence of the perturbation, Eq.~(\ref{trasla}), 
the Lanczos calculation is rather demanding. In fact, the symmetrized Hilbert
space is increased by a factor of four with respect of the standard
calculation on the $J_1{-}J_2$ model,~\cite{poilblanc} and the resulting
dimension of the ground-state subspace is $63.117.760$ for $N=6\times 6$.

\begin{figure}
\vspace{-5cm}
\includegraphics[width=0.50\textwidth]{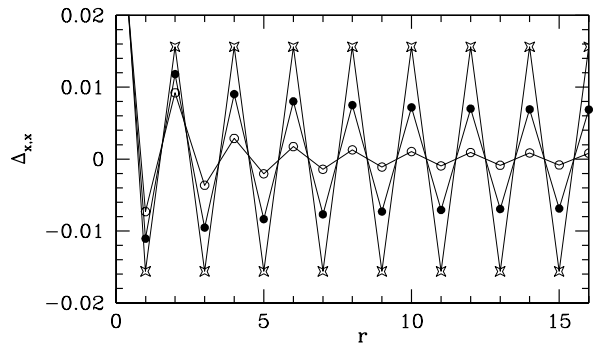}
\caption{\label{dcorr1d}
ED results of the dimer-dimer correlation functions for the 32-site
one-dimensional chain with
$J_2/J_1=0.1$ (empty dots), 0.4 (full dots),
and 0.5 (stars).
Notice however that at the exactly solvable
point $J_2/J_1=0.5$, $|\Delta_{x,x}(r)|=1/64$,
independent on the distance for any non-overlapping singlets.
}
\end{figure}

In Fig.~\ref{pig} we present the ED
results for the susceptibility associated to the operator 
$\hat{O}_{\rm T}$. For comparison, the data for the one-dimensional 
$J_1{-}J_2$ model are also shown.
In the latter case, it is well known that a transition form a 
quasi-ordered state, with power-law spin-spin correlation functions, to a dimerized 
phase occurs at $J_2/J_1\simeq 0.241$.~\cite{emery1d}
The response of the system to the perturbation is very different 
below and above the critical point [Fig.~\ref{pig}(a)-(b)].
This is particularly evident by performing the size-scaling
of $\chi_{\rm T}/N$ which diverges for $J_2/J_1=0.4$,
while saturates to a constant for $J_2/J_1=0.1$ consistently with the known scaling
dimension of the spin-Peierls perturbation in the Luttinger regime, $X=1/2$ [Fig.~\ref{pig}(b)].~\cite{note}
In contrast in the two-dimensional case, for both $J_2/J_1=0.5$ and
$J_2/J_1=0.55$, $\chi_{\rm T}/N$ decreases by increasing the lattice size 
[Fig.~\ref{pig}(c)-(d)] suggesting that a divergence of this
quantity in the thermodynamic limit is unlikely, even if not ruled out
by our small-size calculations.
Indeed, Sushkov {\em et al.}~\cite{russi2} found a divergence of the same
susceptibility $\chi_{\rm T}$, by using a series expansion, that are free from
size effects but are strongly biased by the approximate approach: the limited
number of coefficients known in the infinite series and the difficulty of
reconstructing a possible translational invariant state starting from a
dimerized reference one.
Note that two-dimensional systems behave very differently from 
chains of comparable linear size, where
$\chi_{\rm T}/N$ always increases with $N$ in the dimerized phase
[Fig.~\ref{pig}(a)-(b)]. This points toward the absence of 
dimer order parameter in the strong frustration regime.
   
\begin{figure}
\includegraphics[width=0.45\textwidth]{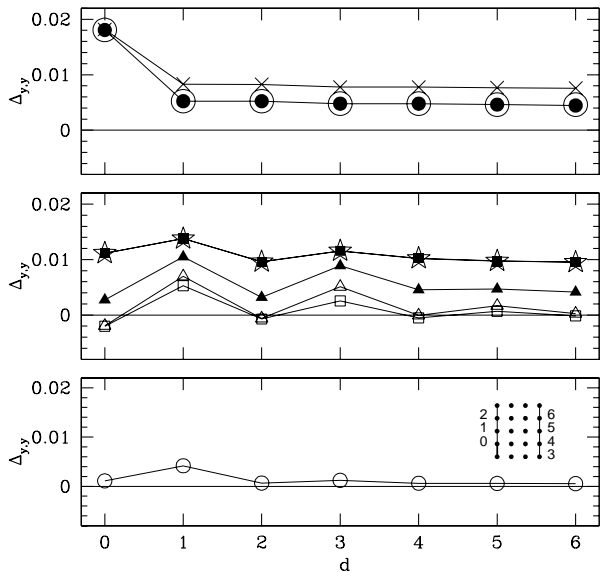}
\caption{\label{dcorr2d}
ED calculation of the dimer-dimer correlations among pairs parallel
to the $y$ axis as a function of the Manhattan distance $d$ for $N=6\times6$.
$J_2/J_1=0.2$ (empty dots),
0.45 (empty squares), 0.55 (empty triangles),
0.6 (full triangles), 0.65 (full squares), and
10 (full dots). The stars (crosses)
indicate the results for the lowest $d$-wave ($s$-wave) excited state for
$J_2/J_1=0.65$ ($J_2/J_1=10$). The large circles in the top panel
are the values for $J_1=0$.
}
\end{figure}

In order to investigate the possible occurrence of a spin-Peierls phase we
now analyze the 
dimer-dimer correlation functions of the unperturbed ($\delta=0$) ground state:
$$
\Delta_{\mu,\nu}(r-r^\prime)
=\langle \hat{S}_r^z\hat{S}_{r+\mu}^z \hat{S}_{r^\prime}^z 
\hat{S}_{r^\prime+\nu}^z\rangle 
- \langle \hat{S}_r^z\hat{S}_{r+\mu}^z \rangle 
\langle \hat{S}_{r^\prime}^z \hat{S}_{r^\prime+\nu}^z\rangle~,
$$
where $\mu$ and $\nu$ are two unit vectors connecting the 
nearest-neighbor sites on the lattice. 
In presence of a broken spatial symmetry, the latter has to converge
to a finite value for large distance, displaying also the typical 
staggered pattern shown in Fig.~\ref{dcorr1d} for the one-dimensional model.~\cite{notess}
In this case, the different decay of the dimer correlations below and above the
transition can be also recognized: oscillatory power-law in the Luttinger regime
and constant amplitude oscillations in the dimerized phase.
In Fig.~\ref{dcorr2d} we present a systematic study of the dimer-dimer
correlations on the $6\times 6$ cluster as a function of $J_2/J_1$.
The lower panel is for $J_2/J_1 = 0.2$ (AF N\'eel phase),
the middle one is for $0.5 \leq J_2/J_1\leq 0.65$ 
and the panel on the top is for $J_2/J_1=10$ (AF collinear phase).

The dimer correlations do not seem to be much affected entering
the non-magnetic region ($J_2/J_1 \gtrsim0.4$) and the large distance 
value of the correlations
for $J_2/J_1 \simeq 0.5$ is comparable to the one in the
AF N\'eel phase, where the dimer correlations
decay to zero with a power law. By increasing further 
the frustration, the dimer-dimer correlations increase 
reaching a maximum around $J_2/J_1 \simeq 0.65$ and then converging
for large $J_2/J_1$ to a non-zero value, which is positive for parallel dimers
and negative for orthogonal ones (not shown). 
Notice, however, that for large enough frustration, $\Delta_{y,y}(r)$ does
not show a staggered behavior, which is inconsistent with the pattern
predicted on the basis of a standard spin-Peierls state.
On the other hand, the  broken translation symmetry of the collinear phase,
also expected in the large-$J_2$ region,
cannot be detected by the dimer-dimer correlation functions plotted in 
Fig.~\ref{dcorr2d}.~\cite{note2}
Instead, the fact that $\Delta_{\mu,\nu}(r)$ has different signs for
parallel and orthogonal dimers is the signature of the breaking of 
the $\pi/2$ rotation symmetry, which also characterizes the AF collinear phase.
In fact, for $J_2\gg J_1$, the system decouples in two unfrustrated Heisenberg models
and $\Delta_{\mu,\nu}(r-r^\prime) \simeq \langle \hat{S}_r^z
\hat{S}_{r^\prime}^z\rangle
\langle \hat{S}_{r+\mu}^z \hat{S}_{r^\prime+\nu}^z\rangle$ 
with the two pairs belonging to
opposite sublattices and $\langle \hat{S}_r^z\hat{S}_{r^\prime}^z\rangle$ 
being the ground-state spin correlations of the 
Heisenberg model on a $N/2$-site cluster 
(see the top panel of Fig.~\ref{dcorr2d}). Hence, in this limit, 
$\Delta_{\mu,\nu}(r-r^\prime) \simeq s_{\mu,\nu} |
\langle \hat{S}_r^z\hat{S}_{r^\prime}^z\rangle 
\langle \hat{S}_{r+\mu}^z \hat{S}_{r^\prime+\nu}^z\rangle|$
with $s_{\mu,\nu}=1$ and $-1$ for parallel and orthogonal dimers,
respectively.
Interestingly enough, for $J_2\gg J_1$, the large distance value of the dimer 
correlations
will approach in the thermodynamic limit $\Delta_{\mu,\nu}(r)\simeq 
(m^\dagger)^4/9$,
where $m^\dagger \simeq 0.307$ is the antiferromagnetic 
order parameter for the spin-half Heisenberg model.~\cite{matteo} 

The non-monotonic behavior of the dimer correlations, displaying a 
maximum around $J_2/J_1 \simeq 0.65$ is likely to be a finite-size effect 
related to the finite-size gap between the two lowest singlet states.
These states cross on the $6\times 6$ for $J_2/J_1 \simeq 0.65$, 
the ground-state symmetry changing from $s$-wave (below) to $d$-wave 
(above).~\cite{poilblanc}
In fact, as we have checked in the AF collinear phase,
the value of the dimer-dimer correlations on the $s$-wave 
singlet (crosses in the top panel of  Fig.~\ref{dcorr2d}) is larger than the value on the $d$-wave one, 
the two values converging to the same 
only when they become degenerate, i.e., at the level crossing on finite-sizes (stars in the middle panel
of Fig.~\ref{dcorr2d}), and in all the AF collinear phase in the thermodynamic limit.

\begin{figure}
\includegraphics[width=0.45\textwidth]{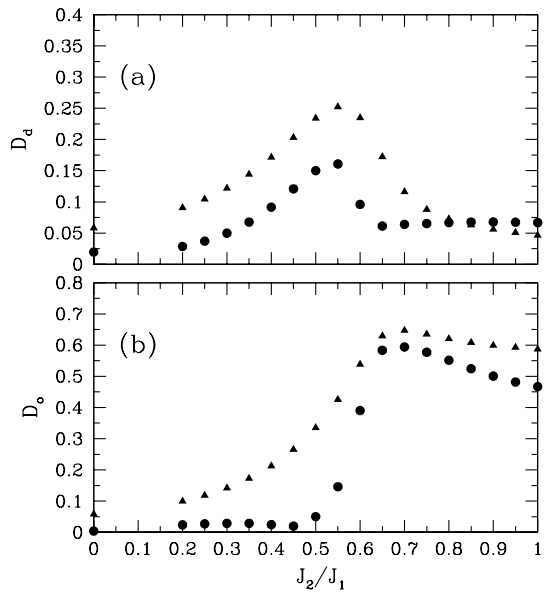}
\caption{\label{ordpar}
(a): ED calculation of the finite-size the dimer order parameter $D_d$ as a function
of $J_2/J_1$, for the $4\times 4$ (full triangles) and the $6\times 6$ 
(full circles) lattice. (b): The same for $D_o$.
}
\end{figure}

Following Ref.~\cite{white}, it is possible to obtain a finite-size estimate 
of the dimer order parameter from the long-distance behavior of 
the dimer-dimer correlations shown in Fig.~\ref{dcorr2d}:
$$
D_d^2=9 \lim_{|r|\to\infty} \vert (\Delta_{y,y}(r-y) - 2\Delta_{y,y}(r) + 
\Delta_{y,y}(r+y)\vert,
$$
where the factor $9$ is required to take into account the three spin
component of the order parameter.
The behavior of $D_d$ as a function of $J_2/J_1$ is reported in
Fig.~\ref{ordpar}(a) for the $4\times 4$ and $6\times 6$ lattices.
A remarkable decrease can be appreciated by increasing the lattice size, 
even in the strongly frustrated regime. 
Moreover, the maximum order parameter on the $6\times 6$ lattice is 
considerably smaller than the corresponding one for the dimerized chain, 
i.e., $D_d=0.75$ for the Majumdar-Ghosh case ($J_2/J_1=0.5$).
These facts should suggest that a sizable dimer order parameter does not 
survive in the thermodynamic limit, leaving the possibility of a disordered 
spin liquid (or, at most, of a ground state with a very small dimer order).

Finally, in analogy with the definition of $D_d$, it is possible to define a dimer
order parameter related to a rotational symmetry breaking:
$$
D_o^2=18 \lim_{|r|\to\infty} \vert (\Delta_{y,y}(r) - \Delta_{y,x}(r) \vert.
$$
Again, the behavior of $D_o$ as a function of $J_2/J_1$ is reported in
Fig.~\ref{ordpar}(b) for the $4\times 4$ and $6\times 6$ lattices.
A very sharp increase of $D_o$ around $J_2/J_1 \sim 0.6$ confirms that a rotational 
symmetry breaking is plausible for large frustration, where the collinear phase is settled
down.~\cite{becca} For smaller values of $J_2/J_1$, instead, the ground state is
rotational invariant, as previously found.~\cite{luca}

In conclusion, we have presented an exact diagonalization study of 
dimer susceptibility and dimer correlations of the spin-half $J_1{-}J_2$ model.
Although we are aware that important finite-size effects occur 
near phase transitions, our analysis does not provide evidence in
favor of a spin-Peierls ground state. By contrast, a homogeneous spin-liquid 
ground state may be realized in the regime of strong frustration.
This possibility has been recently supported by 
a variational calculation showing that a spin-liquid 
projected BCS wave function provides an almost exact 
variational {\em ansatz} of the ground-state in the non-magnetic
phase of this frustrated system.~\cite{caprio2} Work is in progress to 
corroborate these conclusions and to clarify the nature of the excitation 
spectrum of this unconventional state of matter.

We acknowledge useful correspondence with P. Carretta, E. Dagotto, and C. Lhuillier.
We also thank  F. Mila, and D.J. Scalapino for stimulating discussions.
This work has been partially supported by
MIUR (COFIN01) and INFM (PAIS-MALODI). 
L.C. was supported by NSF grant DMR-9817242.



\begin{thebibliography}{99}
\bibitem{note} The scaling dimension $X$ of an operator 
is related to its susceptibility by the relation $\chi\propto N^{2(1-X)}$.
\bibitem{note2} The broken translation symmetry of the collinear phase 
is clearly shown by the peaks of the magnetic structure factor at $Q=(\pi,0)$ and
$Q=(0,\pi)$.
\end{thebibliography}
\end{document}